\documentclass[11pt,twoside]{article}
\usepackage{asp2004}
\usepackage{psfig}
\usepackage{epsf}
\usepackage{graphics}
\usepackage{lscape}
\def\edcomment#1{\iffalse\marginpar{\raggedright\sl#1\/}\else\relax\fi}
\reversemarginpar

\def\Myr{\hbox{\it Myr}}

\def\Lsun{\hbox{\it L$_\odot$}}

\def\Msun{\hbox{{\it M$_\odot$}}}
\def\Minit{\hbox{M$_{\rm initial}$}}

\def\kms{\hbox{km$\,$s$^{-1}$}}

\def\simgr{\mathrel{\hbox{\rlap{\hbox{\lower4pt\hbox{$\sim$}}}\hbox{$>$}}}}
\def\simls{\mathrel{\hbox{\rlap{\hbox{\lower4pt\hbox{$\sim$}}}\hbox{$<$}}}}

\begin{document}

\title{Young Massive Clusters in the Galactic Center}
\author{Donald F. Figer}
\affil{STScI, 3700 San Martin Drive, Baltimore, MD 21218}

\begin{abstract}
The three young clusters in the Galactic Center represent the closest examples
of massive starbursts and they define the upper mass limit of the Galactic cluster mass functions.
In this review, I describe the characteristics and content of the Arches, Quintuplet,
and Central clusters. They each are more massive than any other cluster in the 
Galaxy, and the Arches cluster, in particular, has a mass and age that make
it ideal for studies of massive stellar evolution and dense stellar systems. A
preliminary measurement indicates that the initial mass function in the Galactic
center is top-heavy, suggesting an environmental effect that has otherwise been
absent in similar observations for Galactic clusters. Given the relatively more
evolved nature of the Quintuplet and Central clusters, these clusters contain stars in a wide range of evolutionary
states, including Luminous Blue Variables and Wolf-Rayet stars. The Quintuplet
cluster provides a particularly interesting view of the most massive
stars that are known, the Pistol Star and FMM362. An analysis of the mass spectrum
in the Arches cluster suggests an upper mass cutoff of $\sim$150-200~\Msun.
\end{abstract}

\section{Introduction}
The three young stellar clusters in the Galactic Center are each individually
more massive than any other in the Galaxy. As such,
they represent fertile grounds for exploring a wide variety of astrophysical processes 
over a range of size scales. 
They have enough members to provide good statistics for the high mass range of the initial mass function (10~\Msun$<$\Minit$<$120~\Msun).
They also have coeval
populations with enough mass to populate bins in the initial mass function (IMF) beyond
150~\Msun, a unique property for clusters in the Galaxy. 
They have more massive stars than any other Galactic cluster, allowing one to perform
comparative evolution studies with the assurance that the test points are
all of the same age. Given their ages, the clusters likely contain stars at
all stages of evolution, from the pre-main sequence through end states, including
the Luminous Blue Variable (LBV) and Wolf-Rayet (WR) stages. 
In addition, their unique locale allows one to infer how the initial
mass function might be affected by environmental parameters, i.e.\ cloud temperature. 

Because of their high mass, and apparent top-heavy IMF, the Galactic
Center clusters contain some of the most massive stars in the Galaxy. This is important, as
massive stars are key ingredients and probes of astrophysical phenomena on all size and 
distance scales, from individual star formation sites, such as Orion, to the early Universe 
during the age of reionization when the first stars were born. As ingredients, they control 
the dynamical and chemical evolution of their local environs and individual galaxies through 
their influence on the energetics and composition of the interstellar medium. They likely play 
an important role in the early evolution of the first galaxies, and there is evidence that 
they are the progenitors of the most energetic explosions in the Universe, seen as gamma 
ray bursts. As probes, they define the upper limits of the star formation process and their 
presence likely ends further formation of nearby lower mass stars. They are also prominent 
output products of galactic mergers, starburst galaxies, and active galactic nuclei. 

Despite the importance of massive stars, there is no known firm upper limit to the maximum stellar mass. Such a basic 
quantity escapes both theory, because of the complex interplay between radiation 
pressure and opacity, and observation, because of incompleteness in surveying the 
Galaxy along the plane. The Galactic Center is likely to contain the most massive star
known in the Galaxy, from a statistical perspective, and it does contain several
particularly good candidates. 

In this review, I discuss the properties of the Galactic Center clusters and their
massive stellar content. 

\section{Properties of the Clusters}
Properties of the clusters have been reviewed in \citet{fig99a} and \citet{fig03}, and references therein,
and they are summarized in Table~1. 

\begin{table}[!ht]
\caption{Properties of massive clusters in the Galactic Center\tablenotemark{a}}
\begin{center}
{\footnotesize
\begin{tabular}{lrrrrrrrr}
\tableline
\noalign{\smallskip}
&
Log(M1) &
Log(M2) &
Radius &
Log($\rho1$) &
Log($\rho2$) &
Age &
Log(L) &
Log(Q) \\
Cluster &
\Msun &
\Msun &
pc &
\Msun \, pc$^{-3}$ &
\Msun \, pc$^{-3}$ &
\Myr &
\Lsun & 
s$^{-1}$ \\
\tableline
\noalign{\smallskip}
Quintuplet& 3.0& 3.8& 1.0& 2.4& 3.2& 3$-$6& 7.5 & 50.9 \\
Arches\tablenotemark{b}& 4.1& 4.1& 0.19 & 5.6& 5.6& 2$-$3& 8.0& 51.0 \\
Center\tablenotemark{c}& 3.0& 4.0& 0.23& 4.6& 5.6& 3$-$7& 7.3& 50.5 \\
%NGC 3603& 3.1& 3.7& 0.23& 4.3& 5.0& 2.5& 7.3& 51.1\\
%Trapeziumd& 1.5& & 0.05& 4.7& & 0.3 & $\approx$5\phantom{.} & 48.9 \\
%R136& 3.4& 4.5& 1.6& 2.2& 3.3& $<$1$-$2& $>$7.6& 51.9\\
%small globular (M5)& & 4.8& 4.0& & 2.3& & &  \\
%typical globular(M13)& & 5.5& 3.9& & 3.1& & &  \\
%large globular (M22)& & 6.8& 3.2& & 4.7& & &  \\
%NGC 1705-1&  & 4.9 & 0.9&  & 4.4& 10$-$20& &  \\
%NGC 1569-A& & 5.5& 1.9& & 4.0& 10$-$20& &  \\
\tableline
\noalign{\smallskip}
\tablenotetext{a}
{``M1'' is the total cluster mass in observed stars. ``M2'' is the total cluster mass in all
stars extrapolated down to a lower-mass cutoff of 1 \Msun, assuming a Salpeter IMF slope and an
upper mass cutoff of 120 \Msun (unless otherwise noted)
``Radius'' gives the average projected separation from the centroid position. 
``$\rho1$'' is M1 divided by the volume. ``$\rho2$'' is M2 divided by the volume. In either case, 
this is probably closer to the central density than
the average density because the mass is for the whole cluster while the radius is the
average projected radius. ``Age'' is the assumed age for the cluster. ``Luminosity'' gives
the total measured luminosity for observed stars. ``Q'' is the estimated Lyman continuum
flux emitted by the cluster.}
\tablenotetext{b}{Mass estimates have been made based upon the number of 
stars having \Minit$>$20~\Msun\ given in \citet{fig99b} and the
mass function slope in \citep{sto03}. The age, luminosity and ionizing flux are from \citet{fig02}.}
\tablenotetext{c}{\citet{kra95}. The mass, ``M2'' has been estimated by assuming that a total 10$^{3.5}$
stars have been formed. The age spans a range covering an initial starburst, followed by
an exponential decay in the star formation rate.}
\end{tabular}
}
\end{center}
\end{table}

The three clusters are similar in most respects. 
They each contain $\sim$10$^4$~\Msun\ in stars. 
The new mass estimate for the Arches cluster in the table uses the estimated
number of stars have \Minit$>$20~\Msun\ in the cluster, 160 \citep{fig99b}, and the mass function
in \citep{sto03}, which has a slope of $-$0.9 between 120~\Msun\ and 10~\Msun, and 0
from 10~\Msun\ to 2~\Msun, where the Salpeter value is $-$1.35 \citep{sal55}; 
extending the mass function to lower masses does not add appreciable mass to the cluster. 
The clusters have very high densities, up to nearly 10$^6$~\Msun~pc$^{-3}$, and rival
densities in most globular clusters. They have luminosities of 10$^{7-8}$~\Lsun, and
are responsible for heating nearby molecular clouds. They also 
generate 10$^{50-51}$ ionizing photons per second, enough to account for the nearby
giant HII regions. The primary difference between
the clusters is likely to be age, where the Quintuplet and Central clusters are
about twice the age of the Arches cluster. 

The three Galactic Center clusters define the extreme end in many parameters with
respect to other young clusters in the Galaxy. They are each about a factor
of two more massive than the next most massive young cluster, NGC3603. 
Their luminosities and ionizing fluxes are among the highest in the Galaxy, although both quantities decrease with
age, i.e.\ these quantities for the Arches cluster are a factor of two to three greater
than those for the Quintuplet and Central clusters. It appears that
the the Arches cluster and NGC3603 have similar ionizing fluxes, but the Arches is a factor of two
or three more luminous. 

Given their similar ages and stellar content, the Arches cluster is more similar to NGC3603 than to
any other cluster in the Galaxy. When considering young clusters outside
of the Galaxy, R136 is most similar to the Arches cluster; in this case, the former is probably
a factor of two more massive than the latter. The Arches cluster is a factor of six closer
to us, resulting in a potential advantage in regards to confusion; 
however, it is not observable at UV or visible wavelengths because of the
thick column of dust between us and the cluster.

\section{Properties of the Massive Stars}
The Galactic Center clusters contain a rich set of extraordinarily massive
stars, $\simgr$350 having \Minit$>$20~\Msun. The most massive of these stars 
are in various stages of post main 
sequence evolution, i.e.\ LBV and WR stars. Table~2 gives a summary of the massive
stars in the Galactic Center clusters. The number of O-stars in the case of the
Quintuplet and Central clusters is estimated based upon the ages of the clusters and
the number of identified post-main sequence stars.

\begin{table}[!ht]
\caption{Massive Stars in the Galactic Center Clusters}
\smallskip
\begin{center}
{\footnotesize
\begin{tabular}{lrrrrr}
\tableline
\noalign{\smallskip}
&
O &
LBV &
WN &
WC &
RSG \\
\tableline
\noalign{\smallskip}
Quintuplet & 100 & 2 & 5 & 11\tablenotemark{a} & 1 \\
Arches & 160 & 0 & $\simgr$6 & 0 & 0 \\
Center & 100 & $\simgr$1 & $\simgr$10 & $\simgr$10 & 2  \\
\tableline
\noalign{\smallskip}
Total & 360 & $\simgr$3 & $\simgr$21 & $\simgr$21 & 3 \\
\tableline
\tablenotetext{a}{Includes the Quintuplet Proper Members (QPMs).}
\end{tabular}
}
\end{center}
\end{table}

\subsection{Luminous Blue Variables}

Luminous Blue Variables are characterized by their high luminosities (L$>$10$^6$~\Lsun),
high temperatures (T$>$10000~K), and photometric variability \citep{hum94}. They represent
relatively short phases ($\tau\sim25000$~yr) in the post-main sequence lifetimes of 
massive stars inbetween the O-star and WR phases. 

\citet{fig95} predict that the Pistol Star is extraordinarily massive and 
is surrounded by the largest circumstellar ejecta ever observed (10~\Msun), compared to
a few \Msun\ for $\eta$ Car. Further 
to this claim, \citet{fig98} estimate an initial mass of 200~\Msun, establishing 
the Pistol Star as one of the most massive known. They show that the star is single 
based upon their Keck speckle data and spectra; the former reveal that the star is 
single down to a projected distance of 110~AU (14~mas), while the latter do not 
show an obviously composite spectrum. \citet{fig99c} demonstrate that the 
Pistol Star is indeed the progenitor of its surrounding ejecta which still expands
away from the star at 60~\kms. 

\citet{fig99a} identify a star with spectroscopic and 
photometric properties similar to those of the Pistol Star, and located just a 
few arcminutes away, but still in the Quintuplet cluster. \citet{geb00} 
determine that this star, FMM362, is a ``near twin'' to the Pistol Star, having comparable 
luminosity, and thus mass, and variability. Yet, FMM362 is not surrounded by circumstellar ejecta, 
although we have recently 
obtained near-infrared spectra showing a drastic change in temperature with respect 
to earlier observations. The new spectra do not contain most of the previously 
observed lines, suggesting a much cooler temperature for the observed photosphere. 
While we do not know for sure whether this star is experiencing an eruption, 
the observations are suggestive of such an event. At the very least, the
temperature of the star is highly variable, a characteristic of LBVs as they
transition between quiescent and eruptive stages.

The presence of the Pistol Star and FMM362 in a cluster that is $\sim$4~\Myr\ old is 
a puzzle, given that these stars should not live much longer than $\sim$2~\Myr\ \citep{fig98}.
Note that luminosity is linearly proportional to mass for massive stars (\Minit$\simgr$200~\Msun),
so their lifetimes asymptotically approach 2~\Myr\ with increasing mass \citep{bon84}. 
One solution to the puzzle may be that the stars are binary/multiple, composed of lower mass stars
which have longer lifetimes. Another possibility is that these stars are products of
recent mergers. Indeed, \citet{kim00} simulate the evolution of the Arches cluster, finding
that at least one high mass merger should occur in such a cluster in the first few \Myr\
of its existence. 

IRS16NE is a massive star in the central parsec that has a near-infrared spectrum similar
to those of the Pistol Star and FMM362 \citep{tmb96}. \citet{naj97} estimate a luminosity
and temperature that place the star amongst LBVs in the HR diagram. 
However, the tell-tale variability, characteristic of LBVs, has not yet been
observed for this star \citep{tam96}. Further monitoring might yet
reveal that it is indeed in the LBV stage. If it does not, then it raises the question
of how a star that otherwise appears to be similar to LBVs can resist the instabilities
in such stars. 

See \citet{pau01} for two other potential LBV stars in the central parsec, IRS34W and IRS16C.

\subsection{Wolf-Rayet stars}
All three clusters each contain more WR stars than in other other Galactic cluster. Taken together,
the three clusters contain 10-15\% of all WR stars in the Galaxy. The Central cluster contains approximately 20 WR
stars, with a roughly equal distribution of WC and WN types \citep{kra95,blu95,tmb96,gen03}. The Arches
cluster contains at least half a dozen WNL types \citep{nag95,fig95t,cot95,cot96,blu01,fig02}, but
it contains no WC stars. This is consistent with its age of $\sim$2.5~\Myr\ \citep{fig02}, and
the models in \citet{mey95}.
The Quintuplet cluster contains at least a dozen WR stars, with an equal split between
WN and WC types, excluding the Quintuplet Proper Members \citep{fig95,fig99a,hom03}.

\subsection{The Quintuplet-proper Members (QPMs)}

The Quintuplet-proper members (QPMs) are the five very red sources for which
the cluster was named \citep{nag90,gla90,oku90}.
They are very bright, m$_{\rm K}$ $\approx$ 6 to 9,
and have infrared color temperatures
between $\approx$ 600 to 1,000\, K. 
After dereddening, their integrated infrared luminosities
are in the range 10$^{4.3}$ to 10$^{5.2}$ \Lsun. Oddly, the objects are spectroscopically featureless
at all wavelengths observed, making their spectral classification ambiguous. 

\citet{fig96} and \citet{fig99a} argue that these objects are not protostars, OH/IR stars, or OB stars
still embedded in their natal dust cocoons. 
Instead, they argue that these stars are dust-enshrouded WCL stars (DWCLs), 
similar to other dusty Galactic WC stars \citep{wil87}, i.e.\ WR 140 \citep{mon02} and WR 98A \citep{mon99}. 
\citet{mon01} favor this hypothesis as a result of their analysis of ISO spectroscopy of
the sources. New evidence in support of this hypothesis 
comes in the form of the identification of a carbon feature near 6.2~\micron\ 
in the QPMs' spectra \citep{chi03}, a detection at
x-ray wavelengths \citep{law03}, and a detection at radio wavelengths \citep{lan03}.

If they are DWCLs, then they are dustier than any others, begging the question: Is there
something special about the Galactic Center environment, such as its metallicity, which causes the winds of 
DWCLs to be particularly dusty? If they are not DWCLs, then they represent a new
phenomenon. The same logic applies to the mid-infrared sources in the Central 
Cluster \citep{bec78}. \citet{eis03} show that some of the mid-infrared
sources in the central parsec are indeed DWCLs.

\section{The Slope of the IMF in the Arches cluster}

\citet{mor93} argues that star formation in the Galactic center could favor
high mass stars as a result of environmental conditions, i.e.\ strong tidal 
forces, enhanced cloud turbulence and gas heating, and strong magnetic fields. 
The presumably high metalicity in the Galactic center might also produce a
variation in the spectrum of masses formed there with respect to what is observed
in the disk of the Galaxy, but it is still not clear if the metalicity in the
Galactic center is extrasolar \citep{ram00}.

\citet{fig99b} use HST/NICMOS data to estimate a mass function slope of $-$0.7 for the Arches cluster, 
over a mass range of 6~\Msun\ to 120~\Msun, where
the Salpeter value is $-$1.35 \citep{sal55}. They were unable to estimate a slope for
the Quintuplet cluster, given the degeneracy in the mass-magnitude relationship
for cluster older than 4~\Myr. The same problem exists for the Central cluster. \citet{sto02} 
further refine the estimate for the slope
of the mass function for the Arches cluster, finding a value of $-$0.8, over a mass range of 6~\Msun\ to 65~\Msun. 
The slightly steeper slope with respect to the value in \citet{fig99b} is due to a proper
treatment of differential extinction across the field of the cluster. Both groups note
that significant mass segregration causes a much shallower slope, roughly zero, toward the center
of the cluster. 

\citet{kim00} use N-body simulations of the cluster to determine that the present
distribution of stars in the cluster is consistent with an IMF having 
a slope of $-$0.75. \citet{por02} argue that the present-day mass function is consistent
with an IMF that is similar to the Salpeter value; however, their
analysis requires a cluster mass of 4(10$^4$)~\Msun, or a factor of four above the
observed value.

\section{An Upper Mass Cutoff to the IMF in the Arches cluster}
\citet{fig03} argue that there is evidence of a firm upper mass cutoff to the IMF in
the Arches cluster. 
Assuming an IMF slope of $-$0.9 (see above), we should expect
to see much more massive stars than are currently observed. 
We should expect at least 10 (4) stars more massive than \Minit=300~\Msun, and 
about 30 (11) more massive than \Minit=150~\Msun\ (the numbers in parentheses
are for a Salpeter IMF slope). 
Indeed, we should even expect one star with an initial mass of 1,000~\Msun! 
Yet, we see no stars in the Arches cluster
that are more massive than \Minit=150~\Msun. This apparent deficit might partially be
explained by the short lifetimes of such massive stars. They only live for about 2~\Myr, according
to \citet{bon84}; whereas, the Arches cluster is about 2.5$\pm0.5$~\Myr\ old \citep{fig02}.

\citet{wei04} recently claim a fundamental upper mass cutoff from their analysis
of the mass distribution in R136 in the LMC. They find a cutoff of $\sim$150~\Msun, similar
to the implied value for the Arches cluster. 

\section{Conclusions}
The Galactic center clusters are unique in the Galaxy, providing for a large
range of studies regarding cluster formation, massive stellar formation and evolution,
and feedback mechanisms, to name a few. Given their masses and relative youth, the clusters
contain a large fraction of the massive stars in the Galaxy. Initial esimates indicate
that the IMF in the Galactic center is skewed toward massive
stars. Further, the clusters suggest that there is a firm upper limit to the
most massive star that can form near $\sim$150-200~\Msun. If these measurements 
remain valid, then the IMF is not universal, and there is an upper limit to the maximum
mass of a star. 

\acknowledgements
I thank Richard Larson for interesting conversations regarding the 
upper mass cutoff in the Arches cluster, and for pointing
out the \citet{bon84} reference in regards to the asymptotic behavior of the lifetimes of massive stars.

\end{document}